\documentclass[conference]{IEEEtran}
\IEEEoverridecommandlockouts
\usepackage{cite}
\usepackage{amsmath,amssymb,amsfonts}
\usepackage{algorithmic}
\usepackage{graphicx}
\usepackage{textcomp}
\usepackage{xcolor}
\def\BibTeX{{\rm B\kern-.05em{\sc i\kern-.025em b}\kern-.08em
    T\kern-.1667em\lower.7ex\hbox{E}\kern-.125emX}}
\usepackage[ruled, linesnumbered]{algorithm2e}
\usepackage{algorithm2e,setspace}
\usepackage{booktabs}
\usepackage{multirow}

\usepackage{array}
\newcolumntype{L}[1]{>{\raggedright\let\newline\\\arraybackslash\hspace{0pt}}m{#1}}
\newcolumntype{C}[1]{>{\centering\let\newline\\\arraybackslash\hspace{0pt}}m{#1}}
\newcolumntype{R}[1]{>{\raggedleft\let\newline\\\arraybackslash\hspace{0pt}}m{#1}}

\makeatletter
 \let\old@ps@headings\ps@headings
 \let\old@ps@IEEEtitlepagestyle\ps@IEEEtitlepagestyle
 \def\confheader#1{%
 \def\ps@IEEEtitlepagestyle{%
 \old@ps@IEEEtitlepagestyle%
 \def\@oddhead{\strut\hfill#1\hfill\strut}%
 \def\@evenhead{\strut\hfill#1\hfill\strut}%
 }%
 \ps@headings%
 }
\makeatother

\confheader{
\parbox{18cm}{\centering Accepted copy for Publication at the Design, Automation and Test in Europe (DATE) Conference 2023
\\ 
Final published version available at: https://doi.org/10.23919/DATE56975.2023.10137025}}
    
\begin{document}

\title{Establishing Dynamic Secure Sessions for ECQV Implicit Certificates in Embedded Systems
}

\author{
\IEEEauthorblockN{Fikret Basic, Christian Steger}
\IEEEauthorblockA{\textit{Institute of Technical Informatics} \\
\textit{Graz University of Technology}\\
Graz, Austria \\
\{basic, steger\}@tugraz.at}
\and
\IEEEauthorblockN{Robert Kofler}
\IEEEauthorblockA{\textit{R\&D Battery Management Systems} \\
\textit{NXP Semiconductors Austria GmbH Co \& KG}\\
Gratkorn, Austria \\
robert.kofler@nxp.com}
}

\maketitle

\begin{abstract}
Be it in the IoT or automotive domain, implicit certificates are gaining ever more prominence in constrained embedded devices. They present a resource-efficient security solution against common threat concerns. 
The computational requirements are not the main issue anymore. The focus is now placed on determining a good balance between the provided security level and the derived threat model. A security aspect that often gets overlooked is the establishment of secure communication sessions, as most design solutions are based only on the use of static key derivation, and therefore, lack the perfect forward secrecy. This leaves the transmitted data open for potential future exposures by having keys tied to the certificates rather than the communication sessions. We aim to patch this gap, by presenting a design that utilizes the Station to Station (STS) protocol with implicit certificates. In addition, we propose potential protocol optimization implementation steps and run a comprehensive study on the performance and security level between the proposed design and the state-of-the-art key derivation protocols. In our comparative study, we show that with a slight computational increase of 20\% compared to a static ECDSA key derivation, we are able to mitigate many session-related security vulnerabilities that would otherwise remain open.
\end{abstract}

\begin{IEEEkeywords}
ecqv, implicit, certificate, sts, dynamic, session, key derivation, embedded, security, constrained, automotive.
\end{IEEEkeywords}

\section{Introduction}
\label{sec:intro}
Security is becoming increasingly important in protecting the ever-expanding connections of modern embedded devices. The use of common schemes, e.g., Transport Layer Security (TLS), often proves to be difficult due to the constrained nature of the used devices, which can only allow for a limited performance overhead~\cite{hughes_2022}. In contrast, implicit certificates are showing promise in replacing the traditional security architecture schemes. Implicit certificates offer a lightweight certificate format, and a flexible public key derivation and authentication mechanism that make the use of public key infrastructures more accessible for constrained embedded systems~\cite{pollicino_2020, porambage_2014, sciancalepore_2017, basic_2022, Siddhartha_2020}.   

Different schemes exist based on the implicit certificates, with Elliptic Curve Qu-Vanstone (ECQV) still being the most popular and researched one \cite{campagna_2013}. While there has been numerous research done on ECQV and its use with embedded systems \cite{pollicino_2020, Pullen2019, porambage_2013, porambage_2014, sciancalepore_2017, basic_2022, Siddhartha_2020, zi_yuan_2022}, we noticed that certain security aspects are left out when considering the session key derivation process. The key derivation (KD) and session establishment solutions often neglect a very important key aspect, the \textit{perfect forward secrecy}, specifically, the ephemeral key security characteristic. Forward secrecy allows for a dynamic KD and it considers the state where each newly derived key has a high-enough entropy and is independent of a previous one \cite{whitfield_1992}. This is especially important in session communication, where interactions happen on a frequent basis. We believe that is characteristic often gets neglected due to a believed premise of the necessity for sacrificing the security strength for the performance gain with the limited embedded devices. Rather, what often gets deployed is a static KD where key computations are directly linked to their certificate material. These keys would, hence, only be changed by the change of the certificates and through re-initiating the authentication and session establishment steps. It is, therefore, called a static key exchange, since no other KD function or additional input data is used to mask the present session key which is fully dependent on the current certificate. This can be very problematic in situations where, implementation-wise, either due to the limitations in the system’s architecture, constrained nature of the devices, or neglect from the developers, can lead to longer than the intended use of the same session key. 

Regular key updates are important, as in unfortunate cases, where the session key might get compromised, e.g., via the node capturing attack by compromising a valid device that holds it, all the captured exchanged messages would also be able to be decrypted. Any attack that can compromise the stored device credentials would be able to exploit the statically derived keys. An especially dangerous attack, which is also prevalent in TLS, is the key compromise impersonation (KCI). It is a man-in-the-middle (MitM) attack where an attacker can impersonate the trusted server side to manipulate the key derivation process~\cite{hlauschek_2015}. 
In 2018, OWASP rated for internet of things (IoT) weak, guessable and hard-coded passwords as the number one weakness for the IoT systems, which also considers the key credentials\cite{owasp}.
In fact, based on the study by the SEC Consult between 2015 and 2016, the number of exposed private keys by IoT devices grew by 40\% \cite{sec_consult_2016}.
The ENISA initiative, targeted at investigating automotive security vulnerabilities, listed remote attacks, theft and surveillance as one of the most potent attacks that can happen due to the lack of the required cryptographic functionality support. In their document, all three attacks are affiliated with the lack of forward secrecy for both the wide and local networks~\cite{enisa_2016}. 

To mitigate these security vulnerabilities, we focus on providing a solution that is independent of the rate of the certificate updates, and which ensures that each new communication session would always yield a new key derivation. Additionally, we want to make sure that a session key compromise does not lead to exposure of previous or further keys, i.e., to guarantee perfect forward secrecy. To fulfil these constraints, we present a design based on the Station-to-Station (STS) protocol \cite{whitfield_1992} for a dynamic KD for implicit certificate schemes and extend on the 
general lightweight ECQV implementation by Pollicino et al.~\cite{pollicino_2020}. Furthermore, we investigate the optimization steps for the STS KD protocol execution for the implicit certificates, analyze its applicability for the embedded hardware by implementing and evaluating it on different devices, and compare it with other related implicit certificate schemes. Summarized, our \textbf{main contributions} contained within this work are:
\begin{enumerate}
    \item Design and implementation of a dynamic key derivation approach for implicit certificate architecture schemes using the STS protocol.
    \item Performance and security evaluation of state-of-the-art (SotA) KD implicit certificate schemes by expanding on the existing work from the automotive and IoT domains.
    \item Testing the protocol's feasibility in an automotive system by implementing it on top of 
    a battery management system (BMS) to depict a real-world scenario.
\end{enumerate}

\section{Background on the Security Architecture}
\label{sec:background}
\begin{figure}[!t]
  \centering
  \includegraphics[width=0.80\linewidth]{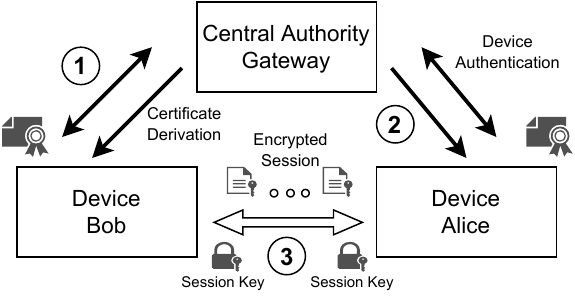}
  \caption{Centralized implicit certificate architecture.}
  \label{fig:general_design}
\end{figure}

We consider three main stages when deploying implicit certificates in a network, as shown in Figure~\ref{fig:general_design} \cite{basic_2022, Pullen2019}: (1) device authentication and deployment, (2) certificate derivation, and (3) session establishment. The deployment phase primarily depends on the main system architecture, however, it generally contains a central, and a more powerful, certificate authority (CA) device. The certificate derivation phase is straightforward with ECQV and almost identical among different 
solutions~\cite{basic_2022, Pullen2019, Siddhartha_2020, porambage_2014}. The session establishment process often differs and depends on the KD and node authentication algorithms. 

\subsection{Key derivation for secure sessions}

We differentiate between two sessions, the certificate session and the communication session. The certificate session considers the validity duration of the currently issued certificates, e.g., in a vehicle during each new engine start, while the communication session considers the duration during one message exchange between two or multiple devices, e.g., monitoring, updates, status readout, etc.

We refer to static key derivation (SKD) as the calculation approach that relies on the traditional Diffie-Hellman KD, i.e., where the keys or the underlying secret are derived from the multiplication of the stored private key and the other device’s public key as $S_k = Prk_a * Puk_b = Prk_b * Puk_a$. The SKD secret is tied to its current certificate session rather than the communication session. As long as the private and public key pairs are not updated, the underlying session key will also not change. Contrarily, the dynamic key derivation (DKD), as the one presented in this work, fulfils the condition that a new session key is derived on each new communication session start, regardless of the current certificate session. The DKD makes sure that each communication session remains independent from the other sessions and should, ideally, provide the perfect forward secrecy attribute. A key derived via this method is also known as the ephemeral secret key.

\section{Related Work}
\label{sec:related_work}

Several research works have already been published on the use of the ECQV and the session KD, both under the general and embedded environments.
Porambage et al.~\cite{porambage_2013, porambage_2014} present one of the earlier session authentication and key exchange solutions for the wireless networks, where the communication between the nodes is done using an SKD. 
For authentication, the protocol uses Message Authentication Code (MAC) with pre-embedded keys, but it also requires that each node possesses from each other the authentication key.
A different authentication scheme is presented by Siddhartha et al.~\cite{Siddhartha_2020}, where an ``authenticator'' is used. It is made out of certificate-related data and signed by the CA. 
A hash function is also used for the additional integrity check. The session key calculation, however, is still based on the standard SKD.

D. Lee and I. Lee \cite{lee_2020} present two approaches to KD in a constrained IoT environment. The first approach is based on the pure ECQV methodology with no additional authentication steps. It relies on validating the identification (ID) and correctness of the certificate calculation, but this does not guarantee the authenticity of the device itself. The certificates and the ID could be spoofed, resulting in a false identification by a malicious actor. Additionally, similar to the work presented by Sciancalepore et al.~\cite{sciancalepore_2017}, the KD uses additional nonces to diversify the key. However, this does not add additional protection since the underlying secret is still calculated using an SKD, i.e., it only considers the multiplication of the private and public keys. The nonces used in the KD can be read from a monitored network. 
Their second method does provide DKD and ephemeral keys, nonetheless, both methods suffer from a central problem, and that is that the device authentication is not considered, rather only the public key validity. 

Recently, Zi-Yuan Liu et al.~\cite{zi_yuan_2022} presented an extension of the ECQV, where devices might house multiple certificates and keys. While novel, the challenges presented in the paper are currently not relevant for this work's use cases, as the focus is placed on larger dynamic networks. 

\section{A Novel Dynamic Key Derivation for ECQV}
\label{sec:design}
\subsection{Security requirements}
\label{sec:sec_requirements}
For the security requirements, it is intended to provide a design that can answer to the following threats: (T1) past data exposure, (T2) MitM attacks, (T3) node capturing attacks, (T4) key data reuse for further session calculations, (T5) key derivation exploitation; each unique key needs to have a high-enough entropy, and that is only stored, and being able to be stored, by the valid parties. We aim to protect two important system assets: session data, and security credentials.
The design also needs to be lightweight in its implementation so as to be easily accessible for the embedded devices.

\subsection{Protocol formalization}
\begin{figure}[!t]
  \centering
  \includegraphics[width=0.52\linewidth]{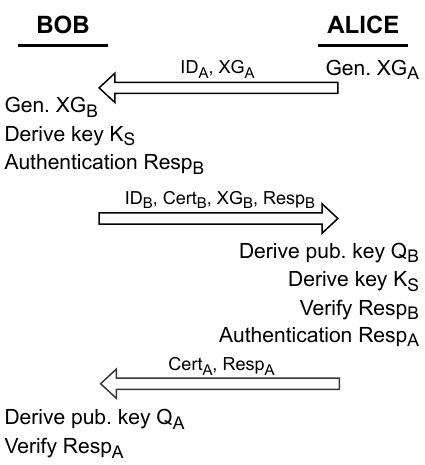}
  \caption{Key derivation using STS protocol for ECQV architectures.}
  \label{fig:sts_prot}
\end{figure}

We base our design of the DKD on the use of the STS protocol \cite{basic_poster_2022, whitfield_1992}. STS is a known protocol used in wide networks; however, it has not been previously investigated for use with the ECQV. The STS derivation should consider the ECQV implicit certificate calculation properties. 
The protocol steps are shown in Figure~\ref{fig:sts_prot}. It is assumed that the first two phases are correctly done as explained in Section~\ref{sec:background}.

The protocol uses the implicit certificate with the elliptic curve digital signature algorithm (ECDSA) to provide authentication as shown with Algorithm~\ref{algo:auth_resp}, and verification with Algorithm~\ref{algo:verif}. What makes it unique compared to other STS algorithm derivations, is that ECQV relies on the implicit derivation of the public key for the signature verification. The security of the ECDSA algorithm with the ECQV scheme has been proven secure against passive attacks \cite{brown_2009}. The public key calculation used for verification is derived as:
\begin{equation}
    Q_{X} = Hash(Cert_{X}) * Decode(Cert_{X}) + Q_{CA}
\end{equation}

The STS provides ephemeral keys, by always deriving a new random elliptic curve (EC) point in the request as:
\begin{equation}
    X\in_{R}[1, ..., n-1] \rightarrow XG = X*G
\end{equation}

Derivation of session $K_S$ keys is done by calculating:
\begin{equation}
    K_{PM} = X_{A} * XG_{B} = X_{B} * XG_{A}
\end{equation}
\begin{equation}
    K_{S} = KDF(K_{PM}, salt)
\end{equation}

\begin{figure}[!t]
  \centering
  \includegraphics[width=0.83\linewidth]{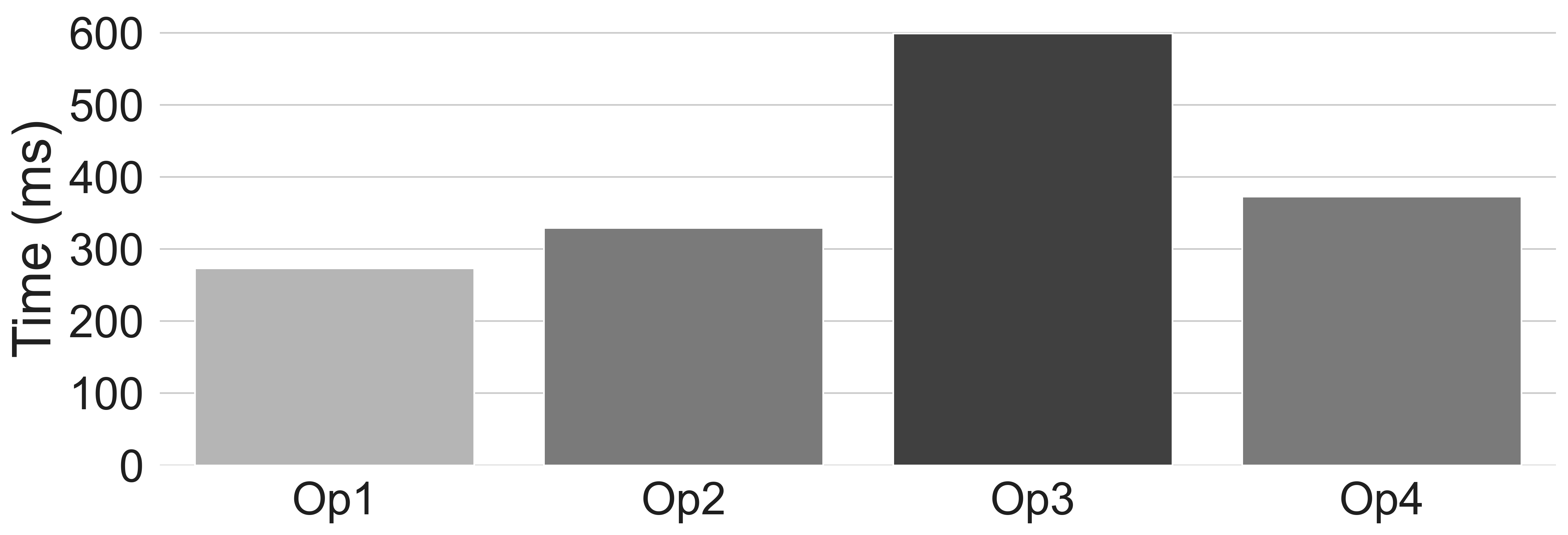}
  \caption{Time duration of individual STS operation runs on an STM32F676.}
  \label{fig:stsopt}
\end{figure}

\begin{algorithm}[!t]
	\caption{STS implicit certificate auth. response.}
	\label{algo:auth_resp}
	\setstretch{0.9}	
	\footnotesize{
		\SetAlgoLined
		\KwIn{\,\,\,\,$XG_{A}$, $XG_{B}$, $K_{S}$}
		\KwOut{$Resp$} 
        \eIf { $ device_A$}{
            $d_{sign} \leftarrow sign(Prk_{A}, (XG_{A}||XG_{B}))$
        }{
            $d_{sign} \leftarrow sign(Prk_{B}, (XG_{B}||XG_{A}))$
        }
		$Resp \leftarrow encrypt(K_{S}, d_{sign})$\\
		\Return{$Resp$}
	}	
\end{algorithm}

\begin{algorithm}[!t]
	\caption{STS implicit certificate sign. verification.}
	\label{algo:verif}
	\setstretch{0.9}	
	\footnotesize{
		\SetAlgoLined
		\KwIn{\,\,\,\,$Resp_{X}$, $Cert_{X}$}
		\KwOut{$Status_{Ok}$, $Status_{Err}$} 
        $d_{signX} \leftarrow decrypt(K_{S}, Resp_{X})$\\
		$Q_{X} \leftarrow hash(Cert_{X})*decode(Cert_{X}) + Q_{CA}$\\
		$Status \leftarrow verify(Q_{X}, d_{signX})$\\
		\Return{$Status$}
	}	
\end{algorithm}

\subsection{STS protocol optimization}
\label{sec:sts_optim}
Even though the STS protocol might provide more security advantages compared to related KD implicit certificate protocols (see Section~\ref{sec:sec_eval}), the main drawback is in its timely execution. As this is still an important aspect of modern constrained systems, we investigate potential optimizations. We divide the entire STS ECQV protocol into four operations:
\begin{itemize}
    \item Op1 - Request phase; random XG point derivation
    \item Op2 - Public key and premaster session key generations
    \item Op3 - Auth. signature derivation and encryption
    \item Op4 - Auth. signature decryption and verification
\end{itemize}

In this analysis, we do not consider the transfer time. We derive two potential optimizations. Similar to the work presented by Sciancalepore et al.~\cite{sciancalepore_2017}, the initial request can be made to contain both the certificate and the XG data, with the calculations of the public key and premaster secret data (see Op2) being done in parallel. Further optimization could be to also include the following Op3 to be executed parallel after Op2 as well. There is a drawback here, and that comes at the expense of the algorithm’s flexibility. Failed authentication requests would only be checked after the calculations have been processed. This could open some doors for misuse by malicious users, either through denial-of-service, or similar attacks. But the actual implementation does not suffer in terms of general security since the calculations are still processed in the same manner. The main advantage would be from the system design perspective, which would allow additional operations to run in parallel.
The sent data is identical to the original protocol, but the message and content order vary slightly. Figure~\ref{fig:stsopt}. shows individual operation time requirements. 

The total execution time with the conventional STS between two devices can be represented as:
\begin{equation}
    \tau_{T} = \sum_{i=1}^{N_{Op}} T_{OpAi} + \sum_{i=1}^{N_{Op}} T_{OpBi}\;\text{ ,with } N_{Op} = 4
\end{equation}

As the optimization can be applied through the Op2 and Op3, we get the following derivation based on the time that each device takes to calculate the operations:
\begin{equation}
    \forall x \in \{2,3\},\,T_{OpAx} =
    \begin{cases}
        0,          & \text{if } A = B\\
        |T_{OpAx} - T_{OpBx}|,& \text{otherwise}
    \end{cases}
\end{equation}

This means that no additional time is taken per device A (or B, as it is symmetrical) if they are identical, or if they are not, the extra amount of time depends on the difference in their execution time for Op2 and Op3.

If the devices are equal, ideally, optimization formulas for two different steps of optimizations based on the system requirements would bring the total run times to:
\begin{equation}
    \text{  Opt. I }\;\tau_{T}^{'} = 2*T_{Op1} + T_{Op2} + 2*T_{Op3} + 2*T_{Op4}
\end{equation}
\begin{equation}
    \text{Opt. II }\;\tau_{T}^{''} = 2*T_{Op1} + T_{Op2} + T_{Op3} + 2*T_{Op4}
\end{equation}

The primary advantage of the optimization is the clear reduction in the total execution time by maintaining a minimal change to the original STS protocol structure. In Section~\ref{sec:protocol_eval}, we compare different protocols for the implicit certificate KD and show the difference in time execution between the optimized and non-optimized STS on real embedded hardware.

\section{Implementation and Evaluation}
\label{sec:evaluation}
\begin{table*}[!t]
\caption{Execution time in milliseconds of the KD protocols for ECQV for the respective embedded hardware.}
\begin{center}
\begin{tabular}{@{}lcccc@{}}
    \toprule
    \textbf{Protocol / Device} & ATMega2560 & S32K144 & STM32F767 & RaspberryPi 4 \\
    \midrule
    S-ECDSA & \multicolumn{1}{C{3.3cm}}{$36859.26\pm0.18$} & \multicolumn{1}{C{3.3cm}}{$2894.1\pm 9.83$} & \multicolumn{1}{C{3.3cm}}{$2521.77\pm5.87$} & \multicolumn{1}{C{3.3cm}}{$18.76\pm0.11$} \\
    S-ECDSA (ext.) & \multicolumn{1}{C{3.3cm}}{$36882.64\pm0.23$} & \multicolumn{1}{C{3.3cm}}{$2976.2\pm11.56$ } & \multicolumn{1}{C{3.3cm}}{$2602.69\pm8.61$} & \multicolumn{1}{C{3.3cm}}{$18.68\pm0.12$} \\
    STS & \multicolumn{1}{C{3.3cm}}{$46262.03\pm0.13$} & \multicolumn{1}{C{3.3cm}}{$3622.71\pm7.034$} & \multicolumn{1}{C{3.3cm}}{$3162.07\pm7.52$} & \multicolumn{1}{C{3.3cm}}{$23.26\pm0.12$} \\
    STS (opt. I) & \multicolumn{1}{C{3.3cm}}{$41680.23\pm1.2$} & \multicolumn{1}{C{3.3cm}}{$3246.55\pm12.97$} & \multicolumn{1}{C{3.3cm}}{$2818.02\pm11.26$} & \multicolumn{1}{C{3.3cm}}{$20.87\pm0.07$} \\
    STS (opt. II) & \multicolumn{1}{C{3.3cm}}{$32410.81\pm1.14$} & \multicolumn{1}{C{3.3cm}}{$2556.84\pm13.13$} & \multicolumn{1}{C{3.3cm}}{$2219.25\pm11.3$} & \multicolumn{1}{C{3.3cm}}{$16.31\pm0.07$} \\
    SCIANC & \multicolumn{1}{C{3.3cm}}{$8990.49\pm0.03$} & \multicolumn{1}{C{3.3cm}}{$721.67\pm0.28$} & \multicolumn{1}{C{3.3cm}}{$628.1\pm0.32$} & \multicolumn{1}{C{3.3cm}}{$4.58\pm0.02$} \\
    PORAMB & \multicolumn{1}{C{3.3cm}}{$17932.17\pm0.05$} & \multicolumn{1}{C{3.3cm}}{$1471.66\pm0.63$} & \multicolumn{1}{C{3.3cm}}{$1263.0\pm0.42$} & \multicolumn{1}{C{3.3cm}}{$8.98\pm0.04$} \\
    \bottomrule
\end{tabular}
\vspace{-1.75mm} 
\end{center}
\label{table:performance_eval}
\end{table*}

\begin{figure}[!t]
  \centering
  \includegraphics[width=0.90\linewidth]{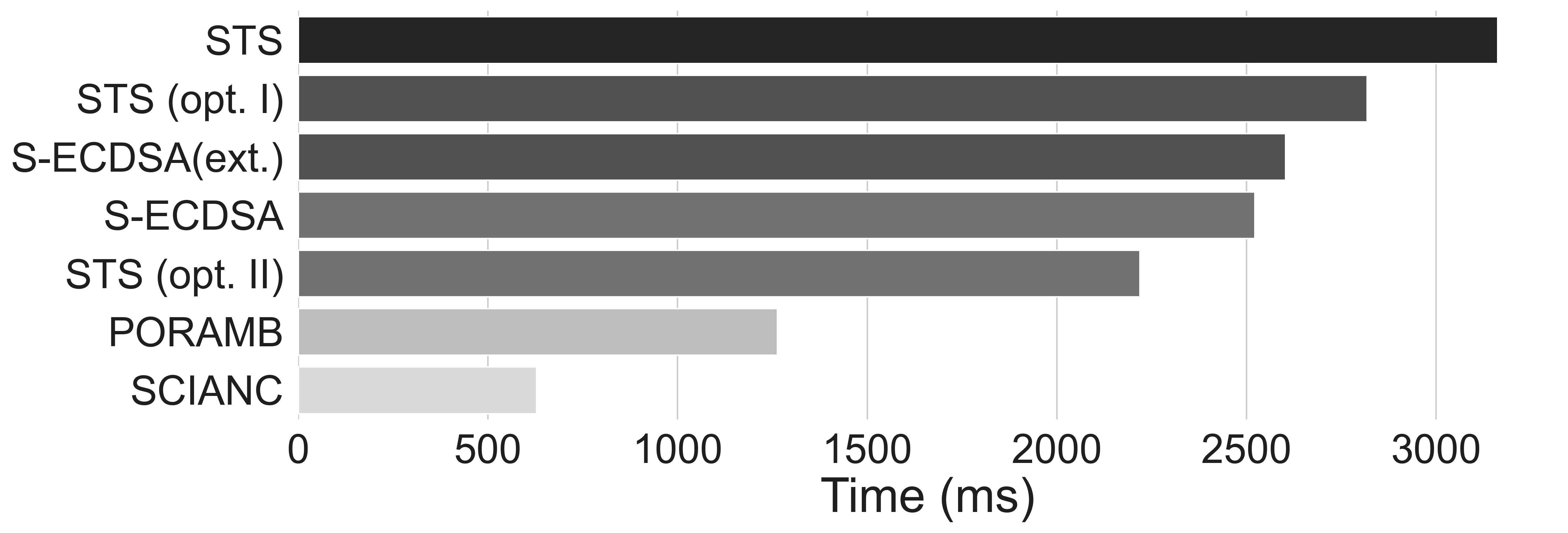}
  \vspace{-2.50mm} 
  \caption{Comparison of the total KD protocols processing time.}
  \label{fig:kd_protcl_times}
  \vspace{-1.75mm} 
\end{figure}

\subsection{Protocol performance evaluation}
\label{sec:protocol_eval}
To show the feasibility of the proposed STS protocol derivation in modern systems and compare it with other SotA KD protocols for implicit certificates, we implement and run the protocols under different embedded devices. We analyze the runs under three main hardware performance level groups:
\begin{itemize}
    \item Low-end: Arduino, ATmega2560, 8-bit 16MHz
    \item Mid-tier: S32K144, ARM Cortex-M4F 32-bit 80MHz; and STM32F767, Cortex-M7 32-bit 216MHz
    \item High-end: Raspberry Pi 4, Cortex-A72 64-bit 1.5GHz
\end{itemize}

The implementations are done in \textit{C} and make use of the functions provided by the micro-ecc, tiny-aes, and bear-ssl libraries, as well as the micro ECQV functions provided by Pollicino et al. \cite{pollicino_2020}. All protocols have been tested with the secp256r1 256-bit EC, with 256-bit level for the SHA and HMAC, and 128-bits for the AES and CMAC.

In total, we test four different protocols derived from two groups based on the use of the authentication mechanism, i.e., on those that rely on the use of ECDSA: (i) static ECDSA by Basic et al.~\cite{basic_2022} as S-ECDSA, and (ii) STS from this work, and those that only use the symmetric cryptography authentication without the EC operations: (iii) from Porambage et al.~\cite{porambage_2014} as PORAMB, and (iv) from Sciancalepore et al.~\cite{sciancalepore_2017} as SCIANC. We also consider the extension of the S-ECDSA protocol, specifically the additional authentication of the ack acknowledgement messages, based on the finished message handling as seen from Porambage et al.~\cite{porambage_2014}. Furthermore, we also evaluate the STS protocol when considering the optimization steps explained in Section~\ref{sec:sts_optim}. Only STS is the true DKD, while the rest fall into the SKD category. The results of the evaluation are shown in Table~\ref{table:performance_eval}, with Figure~\ref{fig:kd_protcl_times}. showing the graphical representation for the STM32F767. The times are averaged after ten runs. The measurements were done using system ticks and Nordic PPK2. The run time scalability is relatively consistent regarding the devices' performances. While STS shows the highest execution time, its optimization variants show the potential time similar to or faster than the S-ECDSA. The PORAMB and SCIANC show the fastest time as they use a different authentication mechanism and do not rely on the EC operations. However, these protocols lack some of the necessary security options as discussed in Section~\ref{sec:sec_eval}. 

\begin{figure}[!t]
  \centering
  \includegraphics[width=0.82\linewidth]{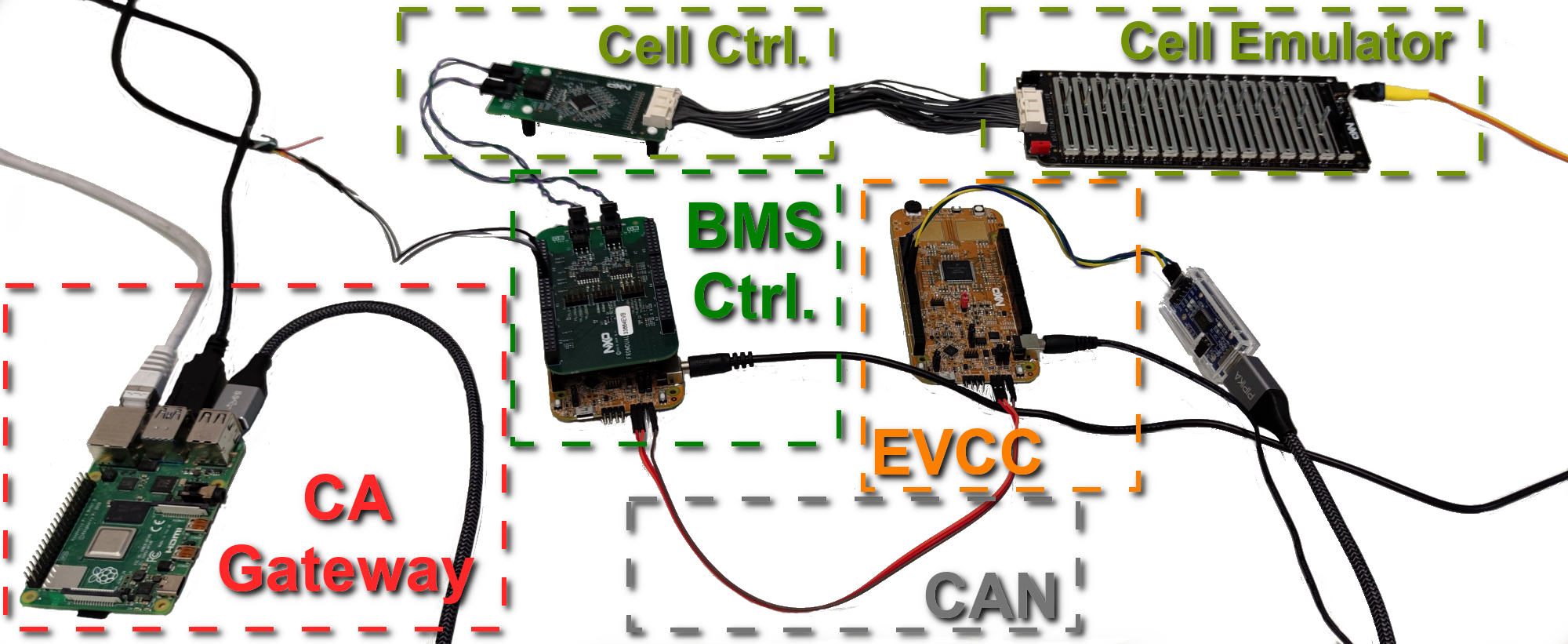}
  \vspace{-2.50mm} 
  \caption{Test suite for the ECQV and KD protocol evaluation.}
  \label{fig:test_suite}
  \vspace{-1.75mm} 
\end{figure}

\begin{figure}[!t]
  \centering
  \includegraphics[width=0.75\linewidth]{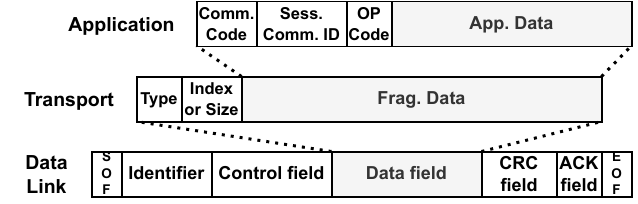}
  \vspace{-2.50mm} 
  \caption{CAN-FD network layers used for the session test communication.}
  \label{fig:can_layers}
  \vspace{-1.75mm} 
\end{figure}

\subsection{Overhead examination}
\label{sec:evaluation_overhead}
To give a clearer analysis of the algorithm processing time, it would be advantageous to consider the transmission overhead, however, that parameter is heavily dependent on the used communication protocol and its configuration. Here, we provide an overview of the overhead for each algorithm during the KD exchange protocol, independent of the communication technology in use. We consider only the protocol-affiliated transmission data on the application level. Security algorithms bit sizes are the same as the ones used in Sect.~\ref{sec:protocol_eval}. We assume IDs to be of 16 bytes and use the minimal certificate encoding with 101 total bytes~\cite{campagna_2013}. The results are shown in Table~\ref{table:overhead_eval}.

\begin{table*}[!t]
\renewcommand{\arraystretch}{1.00}
\setstretch{0.90} 
\caption{Communication steps and transmission overhead of the KD protocols for ECQV.}
\begin{center}
\begin{tabular}{@{}lcccc@{}}
    \textbf{Protocol} & S-ECDSA(+ext.) & STS & SCIANC & PORAMB  \\
    \midrule
    \textbf{Step: Op. (X bytes)} & \multicolumn{1}{L{3.3cm}}{A1: ID(16), Nonce(32)} & \multicolumn{1}{L{3.3cm}}{A1: ID(16), XG(64)} & \multicolumn{1}{L{3.3cm}}{A1: ID(16), Nonce(32), Cert(101)} & \multicolumn{1}{L{3.3cm}}{A1: Hello(32), ID(16)} \\
    & \multicolumn{1}{L{3.3cm}}{B1: ID(16), Cert(101), Sign(64), Nonce(32)} & \multicolumn{1}{L{3.3cm}}{B1: ID(16), Cert(101), XG(64), Resp(64)} & \multicolumn{1}{L{3.3cm}}{B1: ID(16), Nonce(32), Cert(101)} & \multicolumn{1}{L{3.3cm}}{B1: Hello(32), ID(16)} \\
    & \multicolumn{1}{L{3.3cm}}{A2: Cert(101), Sign(64)} & \multicolumn{1}{L{3.3cm}}{A2: Cert(101), Resp(64)} & \multicolumn{1}{L{3.3cm}}{A2: Auth\_MAC(32)} & \multicolumn{1}{L{3.3cm}}{A2: Cert(101), Nonce(32), MAC(32)} \\
    & \multicolumn{1}{L{3.3cm}}{B2: ACK(1), (+Ext\_Fin(96))} & \multicolumn{1}{L{3.3cm}}{B2: ACK(1)} & \multicolumn{1}{L{3.3cm}}{B2: Auth\_MAC(32)} & \multicolumn{1}{L{3.3cm}}{B2: Cert(101), Nonce(32), MAC(32)} \\
    & \multicolumn{1}{L{3.3cm}}{A3: (+Ext\_Fin(96))} &  &  & \multicolumn{1}{L{3.3cm}}{A3 \& B3: Finish(197)} \\
    \midrule
    \textbf{Total} & \multicolumn{1}{L{3.3cm}}{4(+1): 427(+192) B} & \multicolumn{1}{L{3.3cm}}{4: 491 B} & \multicolumn{1}{L{3.3cm}}{4: 362 B} & \multicolumn{1}{L{3.3cm}}{6: 820 B} \\
    \bottomrule
\end{tabular}
\renewcommand{\arraystretch}{1.00}
\setstretch{1.00} 
\vspace{-1.75mm} 
\end{center}
\label{table:overhead_eval}
\end{table*}

Both the S-ECDSA and STS protocols showed similar transmission sizes, with also the least communication steps when not considering the last ack message. The SCIANC protocol also requires only four transmissions, but with only 362 total bytes under the assumed setup. Contrarily, the PORAMB algorithm showed the largest overhead, with 6 total steps and 820 bytes. We did not include the optimized version of STS since it does not differ in terms of the transmitted data. Considering the fast data rates of most communication protocols and the presented data sizes, we can conclude that the influence of the transmission overhead would be minimal in comparison to the individual KD protocols. This is further complemented by the prototype evaluation results from Section~\ref{sec:evaluation_prototype}. 

\subsection{Prototype implementation evaluation}
\label{sec:evaluation_prototype}

\begin{figure}[!t]
  \centering
  \vspace{-1.00mm} 
  \includegraphics[width=0.94\linewidth]{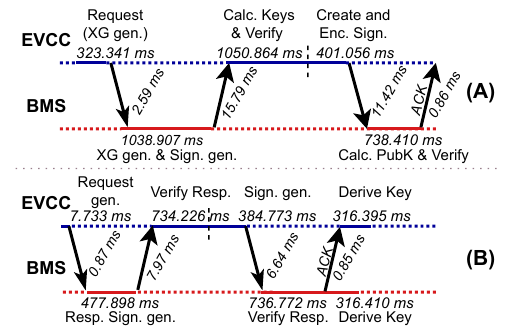}
  \vspace{-2.50mm} 
  \caption{Timeline model of the prototype session communication between a BMS and EVCC for: (A) STS \& (B) S-ECDSA, ECQV KD protocols.}
  \label{fig:timeline_eval}
  \vspace{-1.75mm} 
\end{figure}

In order to evaluate the proposed protocol design on its technical use, we implemented a prototype system that depicts a common communication occurrence between two ECUs in an automotive network. It handles the secure communication between a BMS controller, and an electric vehicle charging controller (EVCC) \cite{fuchs_2020}. Both devices are represented with an S32K144 microcontroller from the NXP Semiconductors to portray a real-world environment. The BMS is additionally connected to a battery cell controller and a battery emulator for emulating a functional unit. The setup is shown in Figure~\ref{fig:test_suite}.

The session communication between the devices takes place over a Controller Area Network (CAN) interface. The test suite uses the CAN-FD derivation with an implemented CAN-TP layer for message fragmentation \cite{ISO15765}. Figure~\ref{fig:can_layers} shows the message formats. 
The devices also communicate with a more powerful CA gateway (represented with a Raspberry Pi 4) to handle the initial device authentication and certificate distribution. The nominal phase CAN-FD bit rate was configured at 0.5\,Mbit/s, with the data phase rate being set at 2\,Mbit/s.

For the evaluation, we compare the proposed STS implementation against the common static ECDSA \cite{pollicino_2020, basic_2022}. For a fair comparison, as to account for the conventional deployment of these protocols in the field, we did not consider the optimization handling for the parallel operation runs argued in Section~\ref{sec:sts_optim}. The implemented security protocols use the same library sources as those mentioned in Section~\ref{sec:protocol_eval}. 
The timeline of both protocols is shown in Figure~\ref{fig:timeline_eval}. The STS implementation showed only a slight difference in the total run time with $3.257\,s$ compared to S-ECDSA's $2.677\,s$,
i.e., an increase of $21.67\,\%$. The CAN-FD transfer time over the physical link was negligible ($<1\,ms$). 
The majority of the communication time from Figure~\ref{fig:timeline_eval}. was for the data processing on the remaining layers.

\subsection{Security analysis}
\label{sec:sec_eval}
We concern ourselves with the listed threats from Section~\ref{sec:sec_requirements} and compare the previous KD algorithms on the provided security level. We also specially look at the mutual authentication procedure, as an important feature against MitM attacks. The analysis is presented in Table~\ref{table:security_eval}, with the following notation: $X$ - weak or no countermeasure, $\Delta$ - partial protection, $\checkmark$ - fully protected.

\begin{table}[!t]
\renewcommand{\arraystretch}{0.85}
\caption{Security overview of the KD protocols for ECQV.}
\vspace{-2.5mm} 
\begin{center}
\begin{tabular}{@{}ccccc@{}}
    \toprule
     & S-ECDSA & STS & SCIANC & PORAMB  \\
    \midrule
    Data exposure & $X$ & $\checkmark$ & $X$ & $X$ \\
    Node capturing & $\Delta$ & $\Delta$ & $X$ & $X$ \\
    Key data reuse & $X$ & $\checkmark$ & $\Delta$ & $X$ \\
    Key der. exploit & $\Delta$ & $\checkmark$ & $\Delta$ & $\Delta$ \\
    Auth. procedure & $\checkmark$ & $\checkmark$ & $\Delta$ & $\Delta$ \\
    \bottomrule
\end{tabular}
\renewcommand{\arraystretch}{1.00}
\vspace{-1.75mm} 
\end{center}
\label{table:security_eval}
\end{table}

\begin{figure}[!t]
  \centering
  \includegraphics[width=0.85\linewidth]{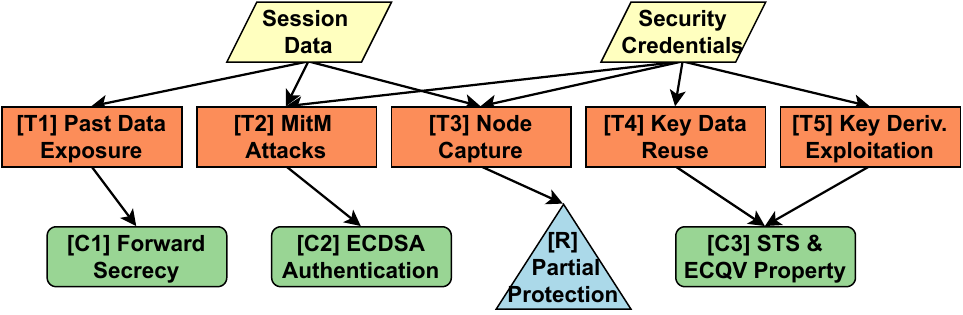}
  \vspace{-2.50mm} 
  \caption{Block diagram representation for the STS-ECQV KD threat model.}
  \label{fig:gsn_model}
  \vspace{-1.75mm} 
\end{figure}

The lack of forward secrecy for all protocols, except STS, makes them highly vulnerable to previous session data exposure, key material reuse (while having the same certificates), and node-capture attacks. However, we note that no algorithm is fully protected against the node-capture attacks, as even with STS, the protection can only be guaranteed for the previous messages, not the future ones.
The mutual authentication for both SCIANC and PORAMB is based on symmetric cryptography with some concerns. PORAMB has the requirement to store individual keys per the number of devices, which makes future updates troublesome. SCIANC algorithm ties its session key with the KD authentication, meaning that if the session key gets exploited so will the future authentication. On the other hand, with S-ECDSA and STS, the authentication is based on the ECDSA with private keys used for signature derivation. Figure~\ref{fig:gsn_model}. shows the derived countermeasures on the listed threats for the STS-ECQV KD.

\section{Conclusion}
\label{sec:conclusion}
In this work, we have presented a key derivation and session establishment model using the STS protocol within the ECQV implicit certificate framework, and its relation and comparison with other KD protocols on embedded devices. While requiring more time, the STS offers a good balance between providing additional security features and certainty without compromising much of the performance. It showed a slight run time increase of $\approx21\%$ compared to a static ECDSA KD protocol, with no additional communication overhead. While other non-EC authentication-based KD protocols showed a noticeable faster execution time, they also lacked the security level acceptable for modern systems. To compensate for the STS run time, we introduced a series of optimization steps for the protocol operations. For future work, we plan to investigate the influence of security modules and hardware accelerators when considering the implicit certificate protocols on embedded devices, especially those related to session establishment.

\section*{Acknowledgment}
This project has received funding from the ``EFREtop: Securely Applied Machine Learning - Battery Management Systems'' (Acronym ``SEAMAL BMS'', FFG Nr. 880564).

\bibliographystyle{ieeetr}
\bibliography{references}

\end{document}